\documentclass[conference]{IEEEtran}
\usepackage[utf8]{inputenc}
\usepackage[inline]{enumitem}
\usepackage{graphicx}
\usepackage{xcolor}
\usepackage{url}
\usepackage{tabularx}
\usepackage{multirow}
\usepackage{colortbl}

\usepackage{booktabs}
\usepackage{xspace}

\newcommand{\model}{our approach\xspace}

\definecolor{maroon}{cmyk}{0,0.87,0.68,0.32}

\begin{document}
\title{Unsupervised Topic Discovery in User Comments}


\author{\IEEEauthorblockN{Christoph Stanik}
\IEEEauthorblockA{
\textit{Universit\"at Hamburg}\\
Hamburg, Germany \\
christoph.stanik@uni-hamburg.de}
\and
\IEEEauthorblockN{Tim Pietz}
\IEEEauthorblockA{
\textit{Universit\"at Hamburg}\\
Hamburg, Germany \\
tim.pietz@uni-hamburg.de}
\and
\IEEEauthorblockN{Walid Maalej}
\IEEEauthorblockA{
\textit{Universit\"at Hamburg}\\
Hamburg, Germany \\
walid.maalej@uni-hamburg.de}
}

\maketitle

\begin{abstract}
On social media platforms like Twitter, users regularly share their opinions and comments with software vendors and service providers.
Popular software products might get thousands of user comments per day. 
Research has shown that such comments contain valuable information for stakeholders, such as feature ideas, problem reports, or support inquiries. 
However, it is hard to manually manage and grasp a large amount of user comments, which can be redundant and of a different quality.
Consequently, researchers suggested automated approaches to extract valuable comments, e.g., through problem report classifiers.
However, these approaches do not aggregate semantically similar comments into specific aspects to provide insights like how often users reported a certain problem.

We introduce an approach for automatically discovering topics composed of semantically similar user comments based on deep bidirectional natural language processing algorithms. 
Stakeholders can use our approach without the need to configure critical parameters like the number of clusters.
We present our approach and report on a rigorous multiple-step empirical evaluation to assess how cohesive and meaningful the resulting clusters are.
Each evaluation step was peer-coded and resulted in inter-coder agreements of up to 98\%, giving us high confidence in the approach.
We also report a thematic analysis on the topics discovered from tweets in the telecommunication domain.
\end{abstract}

\IEEEpeerreviewmaketitle

\section{Introduction} \label{sec:introduction}
\textbf{Context}.
The software market is highly competitive and dynamic. 
For example, the two major app stores, \emph{Apple App Store} and \emph{Google Play Store}, include $\sim$4 million apps \cite{Statista:Online:2019}. 
App vendors release a new version to fix bugs and try out new features \cite{McIlroy2016}. 
App users who are unsatisfied with certain aspects or features are likely to look for alternatives \cite{baltrunas2015frappe, finkelstein2017investigating, williams:2018}. 
This can quickly lead to the fall of even previously popular and successful products~\cite{Li:Satisfaction:2010}.
Therefore, continuously monitoring and understanding users' changing needs and habits is indispensable for the successful survival and evolution of (software) products. 

\textbf{Problem}.
Involving users in the requirements engineering process can help identify users' changing needs and ensure their satisfaction \cite{Kujala:RE:2005}.
To this end, recent research in requirements engineering focused on using explicit user comments like app reviews and tweets \cite{Martin:TSE:2016}.
This field of research analyses written user comments, e.g., to identify bug reports, feature requests, or uninformative text automatically \cite{Chen:2014:AMI, martin2014prediction, Gao:2015:mining, Villarroel:Release:2016, Stanik:Feedback:2019} and to summarize user opinions about the various app features \cite{guzman2014users, Liang:2015, martens:2017}.
Other studies focus on deriving actionable insights for developers to inform future decisions, e.g., on what to release next \cite{Gao:2015, Maalej:Software:2016, Villarroel:Release:2016, Nayebi:2017:release}.

One underlying problem that these approaches help to solve is to cope with the number of user comments app vendors receive.
Research shows that popular apps receive about 4,000 app reviews and 31,000 tweets daily \cite{Pagano:App:2013, guzman2016needle}.
Although approaches like the bug report and feature request classification reduce the number of user comments stakeholders have to look at, the remaining comments are usually still too many to process manually \cite{stanik:diss:2020}.
Related work, therefore, looked into possibilities for identifying topics in, e.g., bug reports to further summarize the user comments.
However, clustering and topic modeling approaches face two major challenges. 
First, they usually have a predefined fixed number of topics or clusters, which are not easy to set and not flexible to change. Second, these approaches have difficulties finding meaningful results on short text documents \cite{quan2010short}.

\textbf{Solution}.
We suggest an automated approach that takes a set of user comments and summarizes it in human-understandable topics.
Our approach needs only a minimal configuration to identify the optimal number of topics automatically.
It relies on SBERT (sentence-BERT) \cite{reimers-2019-sentence-bert}, a BERT variant optimized for generating embeddings.
BERT is a pre-trained language representation utilizing an unsupervised deep bidirectional language system. 
It creates contextualized word embeddings that allow vectors with richer information than other approaches like word2vec's context-free model.
Based on the SBERT model, we create 768-dimensional embeddings of each tweet.
We then reduce the embedding dimensions using UMAP \cite{McInnes2018UMAPUM} to visualize and explore the data, as well as to reduce the input for the clustering technique.
We applied our approach in the telecommunication domain on a Twitter account used by a company to promote products and support their customers.
For this, we developed a rigorous empirical evaluation to quantitatively and qualitatively assess the approach's performance.

\textbf{Results}.
In the first evaluation step (intruder detection), we evaluated the topics' semantic cohesiveness with 200 peer-coding tasks.  
We found that both annotators agreed with the topic model in 95\% of the cases.
In a second step (document to topic assignment), two annotators agreed with the topic model 98.4\% of the time in 2,000 annotation tasks.
In a third step (expert labels), the annotators peer-coded 800 free text annotations to evaluate if both have the same understanding of the topics.
They reached an overall agreement of 91\%.
Finally, we performed a thematic analysis on the dataset to check whether we could interpret the automatically extracted topics.
The results suggest that our approach can produce human-understandable topics.

\textbf{Structure}. 
In Section \ref{sec:approach}, we report on the challenges when developing our approach and state our solution strategies.
Section \ref{sec:study_design} discusses our research questions and the rigorous empirical evaluation design, which can be reused for similar topic discovery tasks.
Section \ref{sec:results} then reports on the results for every evaluation method we applied.
In Section \ref{sec:discussion}, we discuss the implications of our work and present one particular open-source implementation for one specific use case. 
Finally, Section \ref{sec:related_work} summarizes the related work, while Section \ref{sec:conclusion} concludes the paper.
\section{Approach} \label{sec:approach}

\begin{figure}
    \centering
    \includegraphics[width=\linewidth]{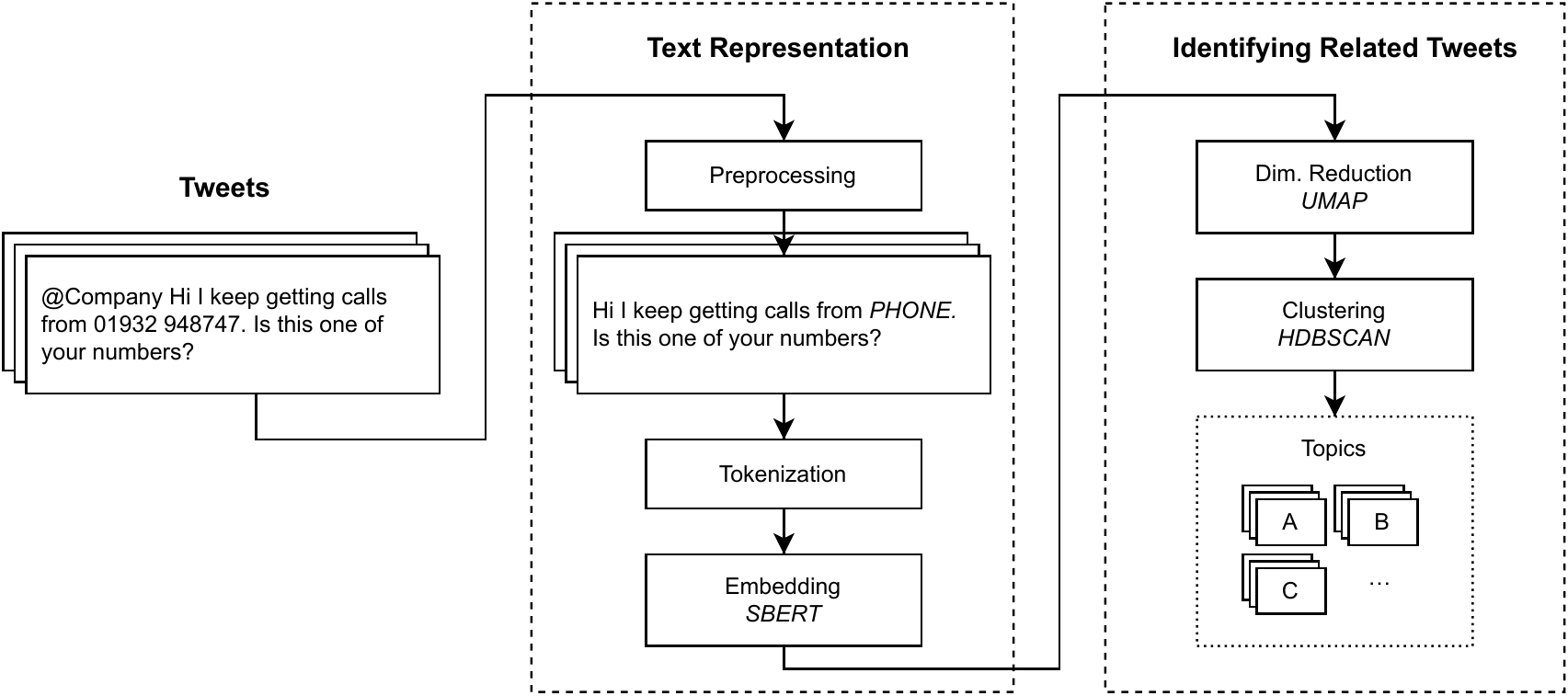}
    \caption{Overview of our approach.}
    \label{fig:approach}
\end{figure}

The main goal of \model is to identify human-understandable topics of semantically similar user comments in tweets.
Figure \ref{fig:approach} shows an overview of the technical steps, separated into two main phases.
The approach takes a set of tweets as input.
It then performs steps to extract the text representation for each tweet.
We create a vector representation also called embedding that is suitable for short text.
Afterward, we use the text representations to group semantically similar tweets into clusters, where each cluster represents one topic.
The following two sections describe the steps for creating the text representation and the clustering in detail.
Our rationale for each individual step is based on a trial-and-error process with a manual inspection of the results.
Finally, we address our reasoning for why we decided to manually inspect the results instead of optimizing based on statistical measures.

\subsection{Text Representation} \label{sec:approach:embedding}
\textbf{Challenges.}
The main challenge is to construct a meaningful embedding for each tweet that captures its semantics and can later be used to find other tweets that discuss similar aspects.
Traditional approaches like \textit{bag of words} or \textit{TF/IDF} are based on word occurrence counts and are therefore neither able to account for semantic relations such as synonyms nor to utilize context information.
However, such information is in particular crucial when working with short texts such as tweets or user comments in general.

\textbf{Approach and Rationale.}
We preprocess all tweets by removing redundant information and specific details that should not be part of the resulting clusters.
As part of this, we remove any mention of the Twitter account that the tweet addresses like ``@MyTwitterAccount`` because when analyzing the tweets of a specific account this part does not offer any discriminative information regarding the tweet's content.
Furthermore, we mask URLs and phone numbers with a single token \textit{URL} or \textit{PHONE} because we have observed that the generated embeddings are sensitive to specific patterns related to this particular data.
For example, we have noticed tweet clusters based on patterns in the digits if not masked.

We then compute a vector representation for the preprocessed tweets to perform computational tasks on them, such as computing their similarity and clustering them.
We decided to use an approach based on BERT \cite{Devlin2019} because of its recent success in various natural language tasks for short texts and tweets specifically \cite{Annamoradnejad2020ColBERTUB, Akiti2020ContextualRO, Roitero2020TwitterGT, Polignano2019AlBERToIB}.
BERT is a deep-learning language model published with multiple model checkpoints trained on large amounts of natural language data \cite{Devlin2019}.
We use SBERT \cite{reimers-2019-sentence-bert}, a BERT variant that is optimized for embedding generation, together with the publicly available SBERT checkpoint \texttt{bert-base-nli-mean-tokens}.
This checkpoint is post-trained on the Natural Language Inference (NLI) and then on the Semantic Textual Similarity (STS) corpus.
With the NLI corpus, the model learns to detect if two sentences are entailments, contradictions, or if they are neutral.
With the STS corpus, the model learns to predict a graded similarity score on a scale from 0 to 5, where a score of 0 indicates that two sentences are completely different while a score of 5 indicates that their meaning is identical.
These datasets allow the model to generate embeddings that are close in terms of their cosine similarity if two sentences are semantically similar.
As both training corpora used to post-train SBERT consist of sentence pair inputs, we also compute embeddings for each sentence of a tweet individually.
We then derive one embedding for each tweet as the average over its sentence embeddings.
We take the average because it matches the inner workings of the \texttt{bert-base-nli-mean-tokens} SBERT checkpoint, which computes the sentence embedding as the average over the token embeddings.
These resulting embeddings for each tweet are 768-dimensional following the size of the hidden vectors of the BERT model.

\subsection{Identifying Related Tweets} \label{sec:approach:cluster}
\textbf{Challenges.}
Identifying semantically similar tweets that mention the same topic based on their embeddings brings along three major challenges.
The first challenge is the computational cost accompanying high-dimensional embedding vectors like the 768 dimensions of the SBERT embeddings.
With some Twitter accounts receiving thousands of tweets per day, possibly accumulating millions in total, the available processing power can become a limiting factor.
The second challenge is that the tweet distribution in the embedding space may vary greatly depending on the Twitter account.
Some possible contributing factors are the total number of tweets, the topical domain, and the scope of interaction with the users.
Therefore, it is unclear whether the formal assumptions of certain clustering methods are met, e.g., the assumption of convex clusters in k-Means.
The third challenge is that some tweets might not be suited to be grouped with other tweets if they are largely unique.
This might, e.g., be the case if the tweet contains significant additional information and should, therefore, be handled separately.
In this case, the tweet should not be assigned to any group with other non-matching tweets.

\textbf{Approach and Rationale.}
To handle the challenge of computational limitations, we first run the dimensionality reduction technique UMAP \cite{McInnes2018UMAPUM} on the embeddings.
Based on preliminary experiments, we chose an output dimensionality of 20, the number of neighbors considered in the algorithm to be 100, and the minimum distance to be 0.
Our preliminary experiments indicated that the results were relatively stable within a certain rate of parameter values, as, for example, changing the dimensionality from 20 to 30 or 15 had only a small impact on the number and size of clusters.

For the clustering step, we chose the HDBSCAN \cite{McInnes2017hdbscanHD} algorithm.
It effectively runs DBSCAN \cite{Ester1996ADA} over multiple neighborhood radii.
We chose HDBSCAN because it does not assume specific cluster shapes and can find clusters with different point densities.
Furthermore, it automatically determines the optimal number of clusters based on the data.
Therefore, HDBSCAN fits well for our use case in which the concrete distribution properties of the data points are unknown in advance.
We set the minimum cluster size to 30, meaning that each generated cluster consists of at least 30 tweets.
We furthermore use the ``leaf cluster'' selection strategy, which prefers smaller clusters.
We chose this strategy to identify fine-grained topics discussed in tweets, such as a network outage in a specific region, as opposed to broader concepts such as ``complaints'', ``requests'', or ``network''.
Lastly, HDBSCAN does not necessarily assign all data points to a cluster---those that do not seem to belong to a particular cluster with sufficient probability will instead be classified as noise.
These noise points correspond to tweets that are not easily grouped with other tweets and therefore should be considered separately.

\subsection{Approach Optimization} \label{sec:approach:optimization}
We base our rationale for the approaches' individual steps on preliminary experiments including a manual inspection instead of optimizing against statistical measures.
For example with HDBSCAN, we manually compared different parameters in terms of selection strategy and minimum cluster size.
We observed the number of topics, their sizes, and contents and compared them based on our goal of a fine-grained topic model and their stability.
An alternative approach would have been to optimize for statistical clustering measures.
Such measures are based on prior assumptions about what a \textit{good} clustering should look like.
These assumptions are intrinsic to the specific evaluation measure.
One example of such a measure is the silhouette coefficient, which is computed on vector spaces like text embeddings.
It yields a value between \(1\) and \(-1\), where a value of \(1\) traditionally indicates a good clustering, \(0\) indicates overlapping clusters, and negative values indicate potentially incorrect clustering.
The assumption behind the silhouette coefficient of what makes up a good clustering is that data points should be similar to all other data points in the same cluster, but dissimilar to all data points of the next nearest cluster.
This assumption naturally favors clusterings that are round, because data points in such clusterings are close to each other and because the different clusters usually do not in any way overlap or intertwine.
Clusterings output by k-Means, for example, optimize for a criterion related to the silhouette coefficient and therefore are likely to achieve good performance when evaluated with the silhouette coefficient.
Without further knowledge about the data characteristics in the embedding space, it is not possible to tell if such assumptions are indicative of a good clustering, particularly in high-dimensional text embedding spaces such as those produced by SBERT.
Both our manual inspection during the approach development, as well as the results of our final evaluation showed promising results.
In contrast, the silhouette coefficient reaches a value of around 0.11 for our final clustering, which would indicate a low-quality clustering and does not correspond to the human annotators' judgments.
\section{Empirical Evaluation Design} \label{sec:study_design}
\subsection{Research Questions} \label{sec:study_design:rqs}
The importance of requirements relevant user comments was highlighted in several scientific publications in recent years \cite{Pagano:App:2013, Martin:TSE:2016, guzman2016needle, stanik:diss:2020}.
Previous work addresses the stakeholders' challenge to filter masses of comments to make sense of them.
Research, therefore, developed approaches that classify user comments as, e.g., problem reports, feature requests, or as irrelevant.
While the results were helpful in practice, stakeholders of enterprise companies or developers of popular apps still receive too many comments for a manual inspection.
Researchers then extended such approaches with, e.g., clustering steps \cite{Villarroel:Release:2016} or the extraction of certain aspects like features \cite{Johann:App:2017}.

However, clustering, in particular, short written user comments, is a technical and empirical challenge.
Technically, it is challenging because clustering approaches rely on embedding text in vector space to calculate vector similarity.
It is further empirically challenging, as the technical approaches rely on individual configurations and fine-tuning.
For example, if we use the topic modeling approach LDA \cite{Blei:2003:LDA:944919.944937}, we have to configure the number of words describing a topic and the number of topics our dataset has.
Additionally, these topics might make sense to the machine, which bases the cluster and topics extractions on its algorithms, but they are not necessarily meaningful or understandable by humans.
Therefore, we have to qualitatively look into the data to find an algorithm and configuration that extracts human-understandable topics.
We thus focus on the following research questions:

\begin{enumerate}[label=\textbf{RQ\arabic*},leftmargin=*]
    \item
        \textbf{How accurately can \model identify human-understandable topics?}
        This research question aims to evaluate the performance of the model which resulted from our approach.
        Here, we performed a multi-step and peer-coded evaluation that focuses on how well humans understand the generated topics.
    
    \item
        \textbf{What topics does \model produce and how actionable are they for requirements engineering?}
        We performed a qualitative exploratory analysis of the clusters to check whether they reveal meaningful topics.
        We further explored the topics' level of granularity.
\end{enumerate}

\subsection{Evaluation Data} \label{sec:study_design:data}
\begin{figure}[tb]
    \centering
    \includegraphics[width=\linewidth]{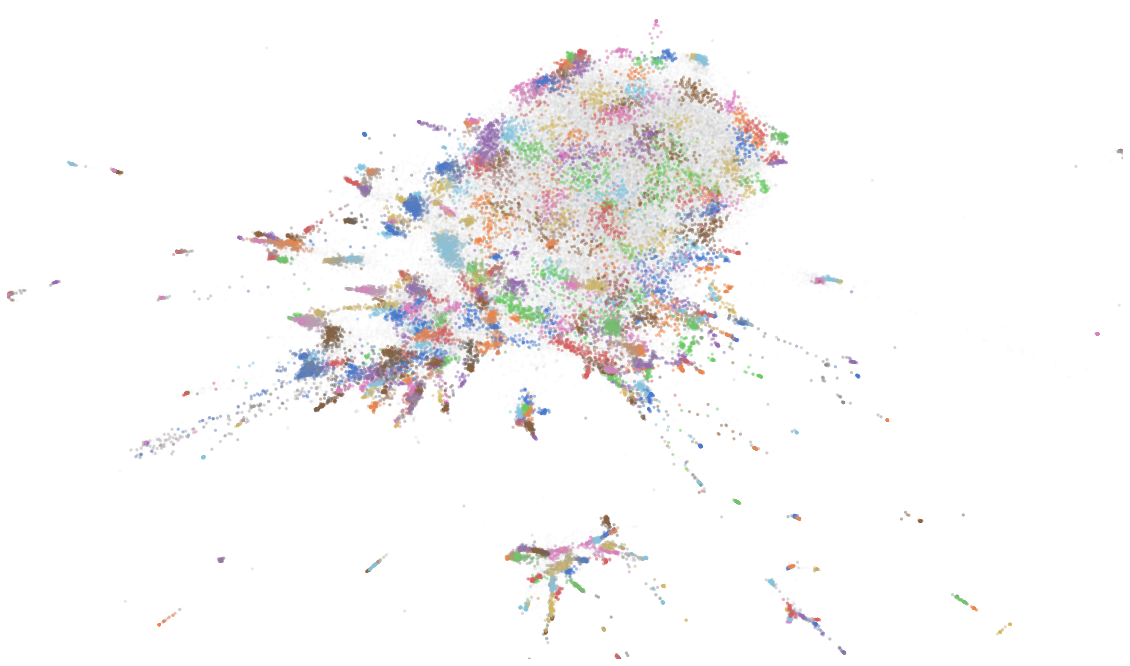}
    \caption{
        Tweet embeddings projected onto two dimensions with UMAP.
        The colored clusters indicate the identified topics.
    }
    \label{fig:embedding_scatter}
\end{figure}

We developed the approach as part of a large multinational research project with a global telecommunication company that aims to better understand their customers' needs.
Their main feedback channel is Twitter, which they are using to promote products, deliver campaigns, and to provide support to their customers.
However, the company's main Twitter channel is not targeting English users.
As the authors are fluent in English and as most software libraries have a well-researched and evaluated support for the English language, we decided to focus on English and report the results here.
Originally, we adapted the approach to their native language.
For the selected case, we crawled 138,639 English tweets for a global telecommunication company.
Figure \ref{fig:embedding_scatter} visualizes the embeddings and identified clusters of all tweets projected onto two dimensions using UMAP.
In total, our approach led to 425 topics covering 23\% of the dataset.
The topics contain 76 tweets on average, with a standard deviation of 65.
We randomly selected a subset for our qualitative analyses.
We describe this in detail in the corresponding sections.

We provide a replication package\footnote{\url{https://mast.informatik.uni-hamburg.de/replication-packages/}} that encompasses the approach's source code, a description of how to use it, as well as the results of all our analyses including the results of each step.
As for the data, we do not have the rights to distribute it publicly but the replication package contains a sample for the approach's input, a link to a library, we used to crawl the data, as well as a contact form for interested researchers.

\subsection{Evaluation Methods} \label{sec:study_design:evaluation}
\begin{figure}[tb]
    \centering
    \includegraphics[width=0.48\textwidth]{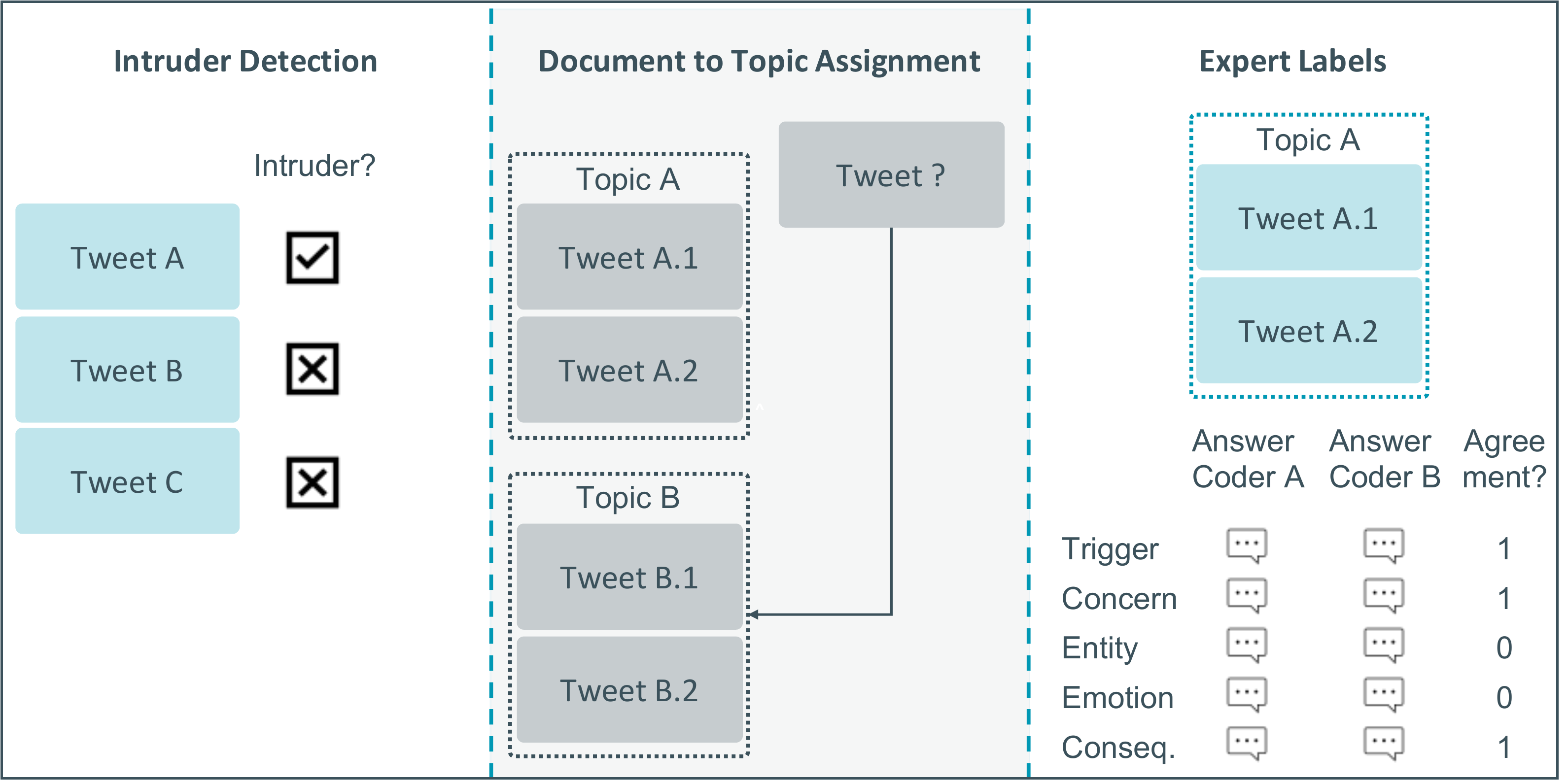}
    \caption{Overview of the three evaluation steps.}
    \label{fig:evaluation_strategies}
\end{figure}
\begin{figure}[tb]
    \centering
    \includegraphics[width=0.48\textwidth]{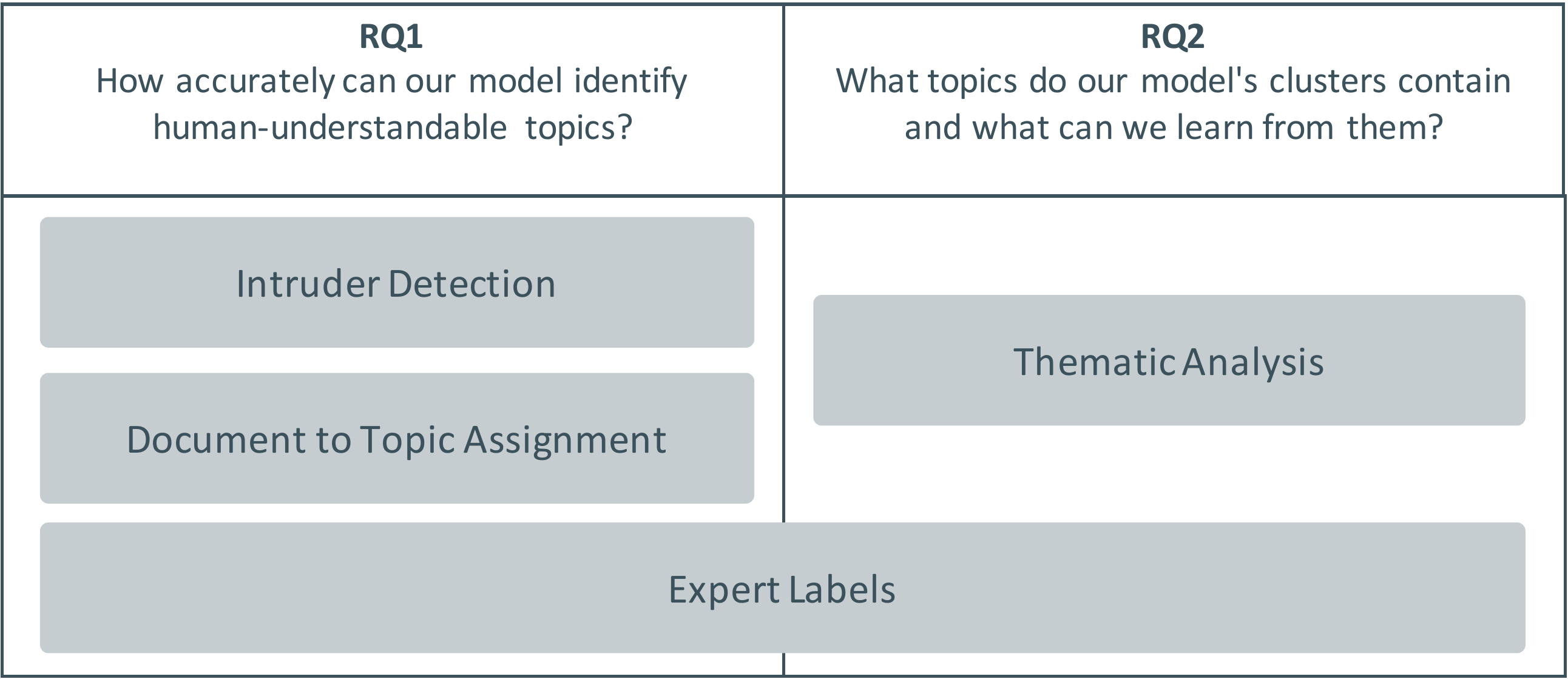}
    \caption{Mapping of our evaluation methods to the research questions.}
    \label{fig:rq_eval_method_mapping}
\end{figure}
We evaluated our approach with a multi-step evaluation, which we summarize in Figure \ref{fig:evaluation_strategies}.
The figure shows that we performed an intruder detection evaluation, a document to topic assignment to check the topics' cohesiveness, as well as a third evaluation step which focused on experts labeling the topics' tweets.
Figure \ref{fig:rq_eval_method_mapping} shows which evaluation method we used to answer the RQs.
In the following, we describe all three evaluation methods in more detail.

\textbf{Intruder Detection}.
Inspired by evaluation methods typically employed in topic modeling, we mirrored the idea from Chang et al. \cite{chang2009reading} who introduced the \emph{word intrusion} and \emph{topic intrusion} method to evaluate our cluster-based topics.
The overall goal is to extend the methodological toolset of statistical evaluation methods that give an idea of the topics' cohesiveness from an algorithmic perspective.
However, that statistical evaluation perspective often differs from a human interpretation \cite{chang2009reading, hall2012evaluating}.
In the original idea, \emph{word intrusion} measures how semantically cohesive the inferred topics are for a human.
In the \emph{word intrusion} method, annotators see a list of words that together build a topic.
However, the sequence of words has one intruder, which the coder must identify.
For example, in the set of words \textit{\{orange, grapefruit, mandarin, lime, cat\}}, the first four words describe fruits, while a cat is an animal.
Therefore, the human task is to identify the word ``cat'' as the intruder.

We designed the first evaluation step similar to the \emph{word intrusion} method.
The goal is to understand whether humans think that tweets belonging to a topic are semantically cohesive.
The left column in Figure \ref{fig:evaluation_strategies} illustrates the idea of this evaluation.
We selected two random tweets from one topic and one random tweet from a different topic.
We then showed the three tweets in random order to two annotators.
The annotators then had to independently select the tweet that semantically does not fit into a topic with the others.
The annotators agree with the topic model if they select the correct intruder, i.e. the one tweet that originated from a different topic than the other two.
We define the label for these three tweets as one coding task.
From all topics our approach extracted, we randomly sampled 200 peer-coding tasks.
We calculated the inter-coder agreement by reporting how often the peer-coders selected the correct intruder.

\textbf{Document to Topic Assignment}.
Chang et al.~\cite{chang2009reading} introduced another evaluation method called \emph{topic intrusion}.
This method measures how well input documents are decomposed into a mixture of topics.
Annotators ``[...]are shown the title and a snippet from a document. Along with the document they are presented with four topics (each topic is represented by the eight highest-probability words within that topic). Three of those topics are the highest probability topics assigned to that document. The remaining intruder topic is chosen randomly from the other low-probability topics in the model.'' \cite{chang2009reading}.

The middle column of Figure \ref{fig:evaluation_strategies} shows how we adapted the \emph{topic intrusion} method to our \emph{document to topic assignment} idea for evaluating our topics.
This is the second evaluation step that shows two independently working annotators two topics, each represented by ten tweets randomly sampled from the topic.
We then randomly sampled ten more tweets total from the two topics.
For each task, we randomized how many of the ten tweets stem from the first and second topic.
We then randomly sampled the tweets from the two topics and shuffled them.
The annotators then had to assign the ten additional tweets to the two topics.
In total, we sampled 200 topic-pairs, each requiring the annotators to assign 10 tweets.
Consequently, we sampled 2000 peer-coding tasks for this evaluation step.
We calculated the inter-coder agreement by reporting how often the peer-coders assigned the tweets to the correct topic.

\textbf{Expert Labels}.
We further developed a third evaluation step to understand if two independent human annotators can interpret the topics and if they share the same understanding of them, which is crucial for learning if the topics contain valuable insights.
If the annotators do not share the same understanding, we do not consider it as an interpretable and insightful topic.

The right column of Figure \ref{fig:evaluation_strategies} illustrates the idea of this evaluation step.
The two annotators see ten randomly sampled tweets for each topic.
Based on the ten tweets of the topic, the two annotators had to answer five questions that describe the topic and its attributes.
The five questions are: 
\begin{itemize}
    \item What event do you think \textbf{triggered} the user's tweet (e.g., a network outage)?
    \item Which \textbf{entities} does the tweet address (e.g., specific teams or companies)?
    \item What is the \textbf{concern} of the tweet (e.g., a complaint)?
    \item How do you judge the \textbf{emotion} of the tweet (e.g., from very negative to very positive)?
    \item Does the user state \textbf{consequences} (e.g., threatening to cancel the contract)?
\end{itemize}
We developed a coding guide in several iterations to help both annotators to answer the five questions accurately.
Then, we created a random sample of 100 topics represented by ten tweets each and let them answer the five questions.
We evaluated the agreement between both annotators by giving either 1 or 0 points for each question---1 if both annotators gave a semantically similar answer and 0 otherwise.
We could then compute an agreement for each coding task (each topic) and compare the overall performance across all coding tasks.

\begin{figure}[tb]
    \centering
    \includegraphics[width=\columnwidth]{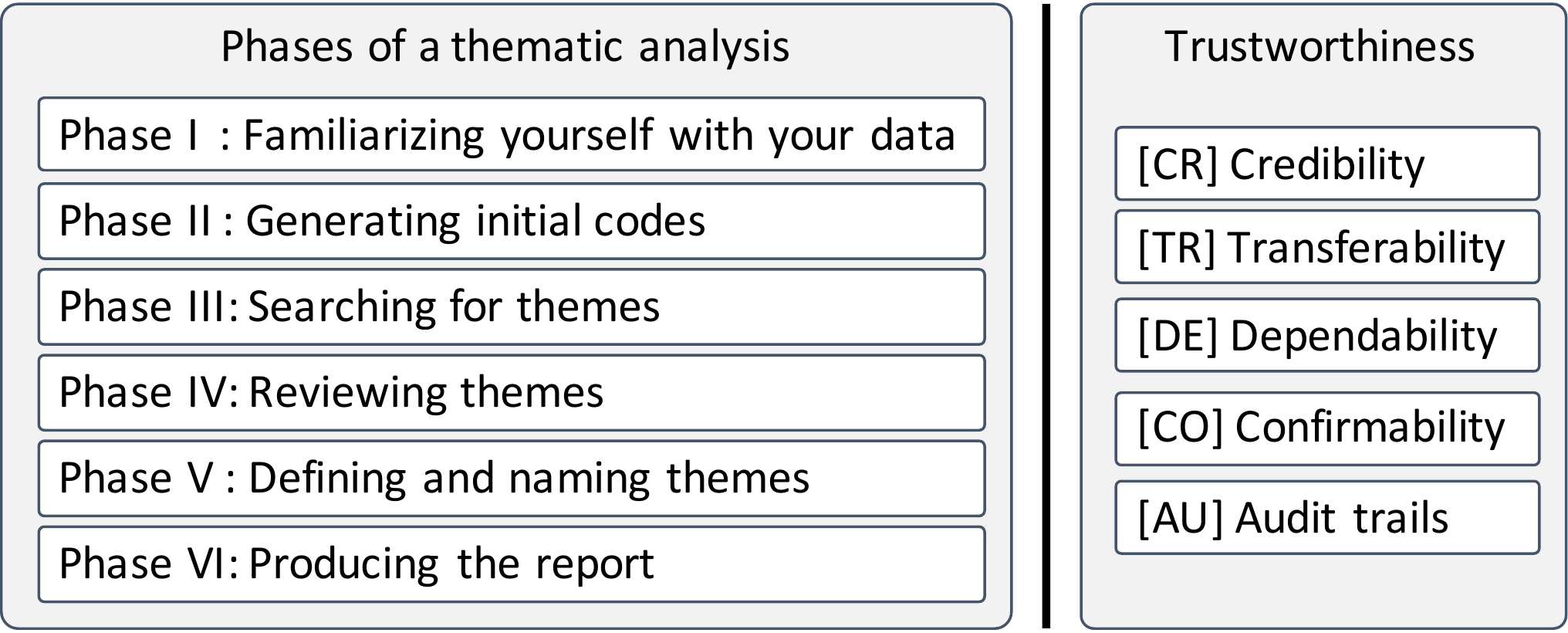}
    \caption{The six phases of a thematic analysis and their criteria to create trustworthiness in the results. Figure adapted from Nowell et al. \cite{nowell2017thematic:trust}}
    \label{fig:thematic_analysis}
\end{figure}

\textbf{Thematic Analysis}.
To answer RQ2, we employed a thematic analysis, in addition to the expert label evaluation.
The goal is to assess if humans have a shared understanding of the topics' content.
For the execution of the thematic analysis, we followed the guideline of Braun and Clarke \cite{braun2006using:thematic:original} and extended it with the trust criteria of Nowell et al.~\cite{nowell2017thematic:trust} whenever we could follow their suggestions.
Figure \ref{fig:thematic_analysis} summarizes the six phases of the thematic analysis we went through and the trustworthiness criteria.
In our following description of how we applied the thematic analysis, we explicitly state the trustworthiness criteria by adding the tags (e.g., [CR] for credibility) to the text as defined in Figure \ref{fig:thematic_analysis}.

We fulfilled phase I by performing the \emph{intruder detection} and the \emph{document to topic assignment} evaluation, for which two authors \textit{[CR]} had to read and work with the data closely.
We further stored the data to make it accessible for other researchers \textit{[AU]}.
We address phase II with the \emph{expert label} evaluation step for which we discussed and decided on the five attributes and questions to answer.
As for all phases, we involved two researchers (annotators) to increase the \textit{[CR]} through researcher triangulation.
We also documented a coding guide \textit{[DE]}, our labels \textit{[AU]} and peer debriefed the phase \textit{[CR]} for creating a shared understanding of the initial labels.
In phase III, the two annotators \textit{[CR]} independently searched for themes in the data and documented their results in tables, which we prepared for the step \textit{[AU]}.
In phase IV, we performed peer-debriefing \textit{[CR]} to discuss the results by comparing the found themes and decided on a final set of themes in phase V, which we also documented in the replication package.
For phase VI, we created one table that merges the independent annotations of the themes \textit{[CR]} and documented the disagreements.
We report on the results of the thematic analysis in Section \ref{sec:results}.
\section{Results} \label{sec:results}




In this section, we present our evaluation results.
We first report on the quantitative results of the three peer-coding tasks to show how accurately our model can identify human-understandable topics.
Afterward, we describe our thematic analysis to summarize what insights we can gather from the topics \model identified in the studied domain.



\subsection{Intruder Detection}\label{sec:intruder_detection}
\begin{figure}
    \centering
    \includegraphics[width=\linewidth]{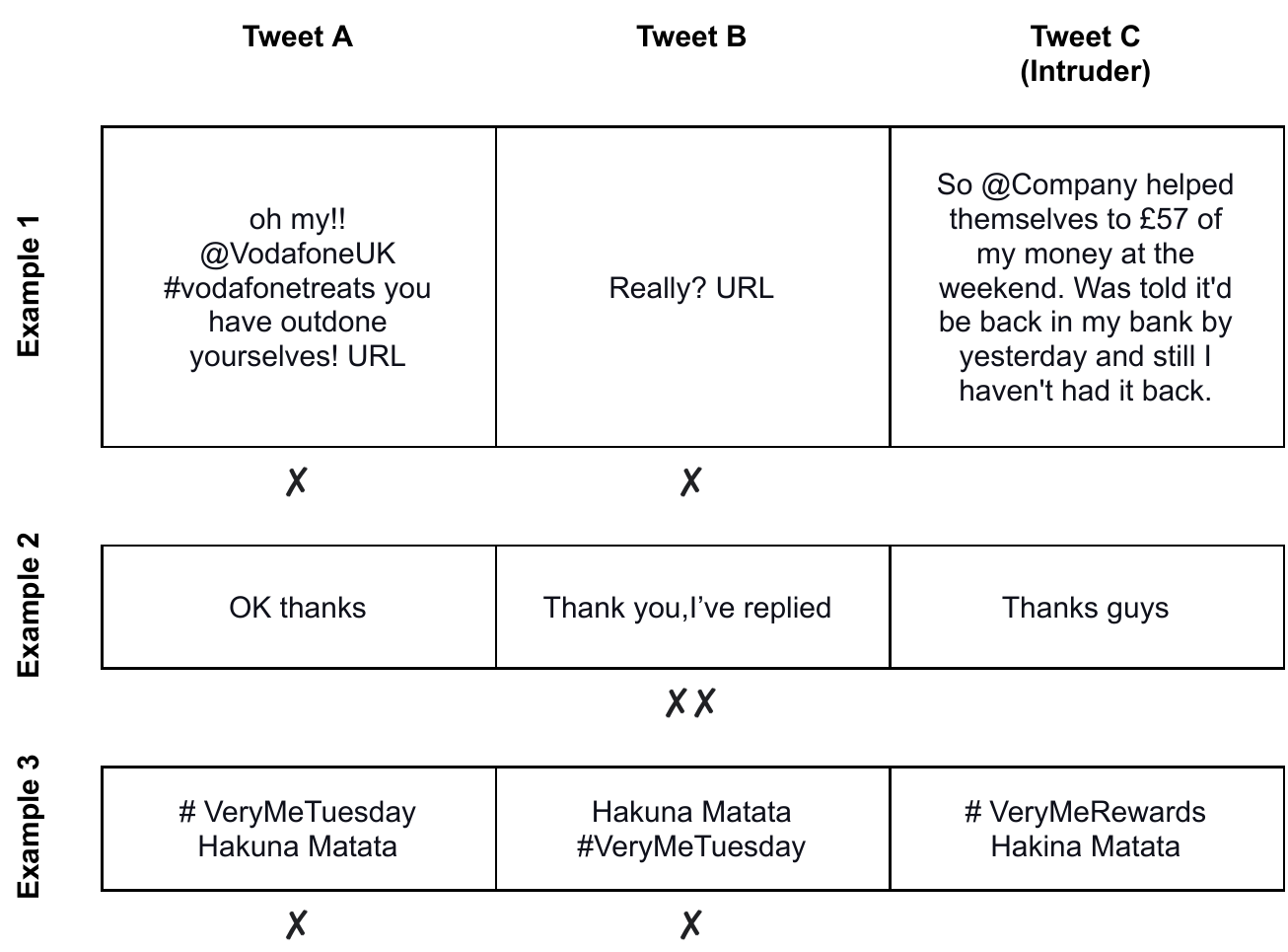}
    \caption{Incorrectly selected by both annotators in the intruder detection task.}
    \label{fig:results:topic_intruders:negative}
\end{figure}

\begin{table}[tb]
\centering
    \footnotesize
    \caption{Intruder Detection and Document to Topic Assignment Results.}

\begin{tabular}{@{}lll@{}}
\toprule
\textbf{}         & \textbf{Intruder Detection} & \textbf{\begin{tabular}[c]{@{}l@{}}Document to\\ Topic Assignment\end{tabular}} \\ \midrule
\rowcolor[HTML]{EFEFEF} 
Peer-Coding Tasks & 200                         & 2000                                                                            \\
Both Correct      & 190 (95.0\%)                & 1968 (98.4\%)                                                                   \\
\rowcolor[HTML]{EFEFEF} 
One Correct       & 7 (3.5\%)                   & 21 (1.1\%)                                                                      \\
Both Incorrect    & 3 (1.5\%)                   & 11 (0.6\%)                                                                      \\ \bottomrule
\end{tabular}
\label{tab:results:topic_intruder_manual_assignment}
\end{table}

In the intruder detection task, the annotators were shown three tweets, two of which the model has assigned to the same topic.
The goal was to identify which of the three tweets does not belong to the same topic.
Table \ref{tab:results:topic_intruder_manual_assignment} summarizes the results of the intruder detection task.
Each coder annotated 200 intruder detection tasks.
Both annotators correctly identified the intruder in 190 tasks, that is 95.0\% of the time.
In 7 tasks, only one of the two annotators correctly identified the intruder.

In the remaining 3 out of 200 intruder detection tasks, neither annotator was able to identify the intruder.
Figure \ref{fig:results:topic_intruders:negative} shows these three cases.
In the first task, the intruder tweet does not include a link while the other two do, albeit in different semantic contexts.
The annotators based their decision on different criteria than \model.
For example, the annotator that chose tweet A as the intruder did so because both B and C might have negative connotations, while A seems more positive.
In the second task, all three tweets have very similar content.
Thus, the annotators were unable to sufficiently differentiate the two topics based on the three tweets alone.
Similarly, the tweets from the third task also seem similar in terms of their content.
In this case, however, only the intruder tweet uses the ``VeryMeRewards'' hashtag, while the other two use ``VeryMeTuesday'', which neither annotator noticed.
As the tweet dataset contains multiple tweets with very similar tweets with either hashtag, two separate topics were created.

The high number of correctly identified intruders and the low number of tasks in which neither annotator found the intruder indicates that the topics found by \model are cohesive with regards to human understanding.

\subsection{Document to Topic Assignment}\label{sec:doc_2_cluster}
In the document to topic assignment task, the annotators were shown sample tweets from two topics.
They were then asked to manually assign further tweets to the two topics.
Table \ref{tab:results:topic_intruder_manual_assignment} summarizes the results of the document to topic assignment task.
Both annotators performed 200 document to topic assignment tasks, each consisting of ten tweets to be assigned.
This results in 2000 peer-coding tasks for the document to topic assignments.
Both annotators assigned the tweets to the correct topic in 98.4\% of the peer-coding tasks.
In 1.1\% of the peer-coding tasks, only one annotator assigned the tweet to the correct topic.

Eleven (0.6\%) of the 2000 tweets were incorrectly assigned by both annotators.
These topics were spread across seven different tasks.
In four of the seven tasks, none of the sample tweets were close to the incorrectly assigned tweets.
One possible reason is that the sample tweets did not sufficiently represent the topics.
We always choose ten random tweets from either topic and show them to the annotators.
Therefore, it is possible that a certain tweet sample does not cover the full semantics of a topic and thus doesn't sufficiently represent it, while another random sample would.
Another possible reason is that the tweet to be assigned does not fit the topic and would therefore be considered as an outlier or false positive member of the topic.
In two other tasks, the annotators could not correctly assign the tweets as the topics' content overlapped.
In one case of these two tasks, the topics were clearly oversegmented as both contained identical tweets but were split into two topics.
In the remaining task, the annotators were unable to sufficiently understand the topics, as the representative tweets did not seem to have any interpretable common topic.

The high number of correctly assigned tweets indicates a natural correspondence between tweets and topics.
The associations between tweets and topics made by \model thus are accurate and intuitive in terms of human understandability.




\subsection{Expert Labels}\label{sec:expert_labels}
In the \textit{expert labels} coding task, the human annotators formulated a short description for each topic based on ten sample tweets.
A topic description is split into the five attributes of
\begin{enumerate*}
    \item the \textit{trigger} for writing the tweet,
    \item their \textit{concern},
    \item the \textit{entity} addressed in the tweets,
    \item the prevailing \textit{emotion} of the tweets.
\end{enumerate*}
In contrast to our evaluation design in Section \ref{sec:study_design}, we do not report on the attribute ``consequence'', as users did not state consequences like canceling the contract in our sample.

Each annotator gave a description for 100 different topics identified by \model.
With the four attributes for each description, this leaves us with a total of 800 peer-coded annotations.
We manually compared the peer-coded annotations for every topic to find out if both annotators agree in their understanding.
We gave each description attribute an agreement score of \(1\) if both annotator's descriptions show that they have the same understanding of the topic regarding the attribute, and a score of \(0\) otherwise.
For the emotion attribute, we gave a score of \(1\) only if both annotators selected the same emotion from the 5-point scale.
As the other attributes allowed free-text answers, we gave an agreement score of \(1\) if both annotators gave a semantically similar attribute description.

To define what counts as ``semantically similar,'' we first inspected a small subset of expert label annotations, and formulated the following rules based on our acquired insights:
\begin{enumerate}[label=R\arabic*]
    \item
        The agreement score should not be influenced by differences in wording. \\
        \textbf{Example:}
        We consider ``Network status inquiry'' and ``Details about network problem'' semantically similar. \\
        \textbf{Rationale:}
        Both descriptions refer to the same underlying concern, as they describe tweets that ask for more information about the state of the network.
    
    \item
        If one annotator found an attribute important for the topic while the other did not, we count this as a disagreement. \\
        \textbf{Example:}
        If one annotator enters ``Buy, Trade-In'' as the concern while the other notes that the topic does not have a specific concern, it results in a disagreement. \\
        \textbf{Rationale:}
        The annotator that does not specify an attribute description implies a broader topic than the other annotator.
        For example, in a topic with a specific device such as the ``Huawei P30'' as the entity, a concern of ``Buy, Trade-In'' would restrict the topic further compared to a topic without a concern, which could also include tweets about problems, returns, or questions.
        As the topics described by the annotators are therefore different, we count it as a disagreement.
    
    \item
        If the annotators choose different levels of abstraction, it should count as a disagreement \\
        \textbf{Example:}
        ``Contract'' and ``Payment'' are disagreements. \\
        \textbf{Rationale:}
        Similar to R2, such differences indicate that the annotators have a different understanding of the topic granularity.
        While a topic of tweets concerning contracts, in general, may include inquiries about payments, it could also include other tweets about terms and conditions or contract changes.
        We therefore count it as a disagreement.
    
    \item \label{itm:expert-labels:swap}
        If the annotators do identify the same concepts as important for the topic, but name them under different description attributes, they should still count as an agreement. \\
        \textbf{Example:}
        A telecommunication company did a raffle campaign under the hashtag ``\#GalaxyA20e'', asking the participants to name their favorite feature of the Samsung Galaxy A20e smartphone.
        For a topic related to this campaign, one annotator identified the trigger as ``\#GalaxyA20e'' and the entity as the ``Fingerprint scanner''.
        The other annotator specified the trigger as ``question from the Twitter account'' and the entities as ``GalaxyA20e, Fingerprint''.
        In this example, we treat both the trigger as well as the entity as agreements. \\
        \textbf{Rationale:}
        In the example, the annotations for neither attribute matches completely.
        The first annotator named the ``\#GalaxyA20e'' hashtag in the trigger, while the other labeled the smartphone as an entity.
        Thus, both annotators specify the same core concepts, which indicates that they share a similar understanding of the topic.
        As the goal of this task is to evaluate the topic understandability, we decided to treat such annotations as agreements.
\end{enumerate}

\begin{table}[tb]
\centering
    \footnotesize
    \caption{Per-Attribute Agreement in the Expert Labels Task}
\begin{tabular}{@{}lrrr@{}}
\toprule
\textbf{Attribute} & \textbf{Annotations} & \textbf{Non-Null Annotations} & \textbf{Agreement} \\ \midrule
\rowcolor[HTML]{EFEFEF} 
Trigger            & 200                  & 177 (89\%)                    & 97\%               \\
Concern            & 200                  & 145 (73\%)                    & 87\%               \\
\rowcolor[HTML]{EFEFEF} 
Entity             & 200                  & 122 (61\%)                    & 97\%               \\
Emotion            & 200                  & 195 (98\%)                    & 84\%               \\ \midrule
\rowcolor[HTML]{EFEFEF} 
\textbf{Total}     & \textbf{800}                  & \textbf{639 (80\%)}           & \textbf{91\%}      \\ \bottomrule
\end{tabular}
\label{tab:results:expert_labels:attribute_results}
\end{table}

Table \ref{tab:results:expert_labels:attribute_results} outlines the results of the expert label task.
The annotators specified a description attribute value for 639 (80\%) of the 800 peer-coded annotations.
More specifically, they specified an emotion in 98\%, a trigger in 89\%, a concern in 73\%, and an entity in 61\% of the topics.
For the remaining 161 (20\%) cases, the annotators did not specify any description attribute as they were not part of the topic, e.g., the topics did not address a particular entity.
The annotators agreed in their understanding of the topic 91\% of the time.
They most often agreed on the trigger and the entity mentioned in the tweets, both showing an agreement of 97\%.
They agreed on the concern for 87\%, and on the emotion for 84\% of the topics.
We noticed that the annotators specified the same concepts but in different attributes (corresponding to evaluation rule \ref{itm:expert-labels:swap}) in 19 cases.

In another 12 of the 800 peer-coded annotations, we noticed that the annotators missed writing down a value for attributes in cases where it seems obvious from the sample tweets.
For example, in a topic of tweets related to the 5G network expansion, one annotator forgot to write down the network provider itself as the entity.
In another case of a topic in which every single of the example tweets is a question, one of the annotators forgot to specify that the trigger for the tweets is that its authors had a question.
For cases such as this, where the example tweets make it immediately obvious that one of the annotators simply forgot to write it down, we decided to count them as agreements.
Our rationale is that such disagreements do not stem from differences in topic understanding but from human annotator limitations.
Counting them as disagreements would not contribute to our goal to measure the topic understandability.

With regards to RQ1, the 91\% annotation agreement show that both annotators were able to reach a common understanding of the topics in most cases.
With regards to RQ2, we were able to learn that a topic nearly always captures tweets with one specific emotion.
We also found that the authors of tweets for a given topic most of the time had the same trigger for writing their tweets or showed the same concern.
Only about half of the tweets, the tweets addressed specific entities.
\vspace{-.025cm}

\subsection{Thematic Analysis}\label{sec:thematic_analysis}
With our thematic analysis, we aim to understand which kinds of topics \model could identify from informal comments as the Tweets.
For this, two annotators first independently defined an initial set of themes based on the peer-coded results of the expert labels task.
Our analysis of the initial labels showed that the annotators often chose different levels of granularity for the themes.
One annotator, for example, decided to separate tweets related to bills, payment, and contracts into different themes, while the other annotator used one single theme for all of them.
We reviewed all initial themes proposed by the annotators and decided on a final common set of themes.
These themes include:
\begin{enumerate*}[label=\alph*)]
    \item tweets about the provider itself,
    \item tweets related to service statues, such as of the network, website, or app, and
    \item tweets about devices.
\end{enumerate*}
Furthermore, we also defined sub-themes for some of the themes, to allow for a more fine-grained analysis of the results later.
For the \textit{devices} theme, for example, we defined sub-themes for smartphones, smartwatches, headphones, and others.
We specified 11 different themes and furthermore defined 12 sub-themes spread across 4 themes.
After re-annotating the dataset with the updated set of themes, the annotators agreed on the themes for 85\% of the topics generated by \model.
We then manually resolved all disagreements through discussion.

Our analysis of the resolved theme annotations showed that 25\% of the topics are about the network provider itself.
10\% of the topics comprise tweets specifically about the mobile or broadband network.
Further 6\% address the status of provided services by mentioning or asking about outages, downtimes, or the speed of the providers' network, website, and app.
Around 18\% of the topics relate to customer support requests, with questions e.g. about buying devices or trade-ins, or follow-ups like asking the support if they have received the customer's private message.
Approximately one in three support request topics consist of tweets that pose a question, the remaining two being follow-up tweets.
Another 18\% are about devices, mostly (14\%) about smartphones.
A large portion of the topics relate to PR campaigns of the network provider.
15\% are replies to raffles that offer participants a chance to win a specific device, e.g., a smartphone.
Similarly, 34\% are topics of users participating in campaigns that offer mostly other kinds of rewards such as additional mobile data volume or gift vouchers.
\section{Discussion} \label{sec:discussion}
We summarize the findings and implications of our work, discuss potential applications, as well as the threats to validity.

The main goal of this work is to identify human-understandable topics within tweets.
We employed a context-sensitive embedding technique called \emph{SBERT} that uses a bidirectional language representation to achieve this goal.
Context-sensitive embeddings include the surrounding tokens for any word and contain more information than context-free embedding techniques like TF/IDF or Glove do.
Therefore, we could achieve coherent topics on the short texts of tweets.

\textbf{RQ1.}
With RQ1, we aimed to understand how accurate our approach can extract human-understandable topics.
For that, we performed an \emph{intruder detection}, \emph{document to topic assignment}, and an \emph{expert label} evaluation.
The \emph{intruder detection} evaluates if the topics are semantically cohesive.
After peer-coding 200 tasks, we found both annotators agreed with the topic model in 95\% of the cases, while there were three tasks in which neither of the annotators could correctly identify the intruder.
In the \emph{document to topic assignment}, we evaluated whether humans can correctly assign a single tweet to a topic if they get two topics to choose from.
Therefore, we can understand if a human can distinguish the topics by reading their tweets and select the topic it belongs to if given a new tweet.
For this evaluation, our annotators peer-coded 2,000 tasks.
They achieved a correct assignment in more than 98\% of the tasks.
We then checked our understanding of the topics with the \emph{expert label} evaluation by analyzing the content in more detail.
With 800 peer-coded free text annotations for the four categories ``trigger'', ``concern'', ``entity'', ``emotion'', we evaluated if we share the same understanding of the topics.
The annotators mean agreement is 91\% across the categories.

\emph{Consequently, we are confident to conclude RQ1 with the result that humans can understand the extracted topics to a high degree as indicated by the annotation agreements of the three qualitative evaluation methods.}
\\
\\
\textbf{RQ2.}
As the \emph{expert label} evaluation is about extracting the tweet topics' content, it also part of answering RQ2, which aims to understand the discovered topics.
By annotating the data following the categories like \emph{trigger} and \emph{emotion} defined Section \ref{sec:study_design}, we extracted the key elements from the topics that enable a human understanding.
We further performed a thematic analysis on top of the expert label results to gain a deeper understanding of the topics by extracting the themes and sub-themes.
In total, we defined eleven themes in the data.
While one theme is for topics we could not understand (``unknown''), and another is for content that does not belong to any of the other defined themes (``other'').
Therefore, we have nine themes that describe the content, such as ``requests'', ``network'', and ``device''.
Besides these top-level themes, we identified 12 sub-themes.
For example, users may address different devices (top-level theme) such as smartphones, smartwatches, or headphones (sub-themes).
Most topics either address campaigns (33\%) like raffles, address the company directly (17\%) or request more information (12\%).

\emph{As a result we found that our approach creates topics that can distinguish the sub-themes like the specific campaigns.}

\subsection{Field of Application} \label{sec:discussion:application}
Several studies from recent years \cite{pagano2013user, Villarroel:Release:2016, di2016would, Maalej:REJ:2016, di2017surf} performed interviews, case studies, and surveys with practitioners and evidently reported the importance of written user comments in practitioners' workflows.
These studies further highlight that practitioners need automated support for aggregating written user comments to overcome the often massive amount of received comments.
Since bare research algorithms are not easily accessible for practitioners, research suggested \cite{Maalej:REJ:2016} and contributed with open source prototypes such as \textit{SUPERSEDE} \footnote{\url{https://github.com/supersede-project}} and \textit{OPENREQ}\footnote{\url{https://github.com/openreqeu}} that help with the problem at hand \cite{stanik:diss:2020}.

Unsupervised approaches allow practitioners to explore data to find prevalent and emerging patterns which supervised classification approaches cannot find as we have to first train them for any information we are looking for in the data. 
Therefore, unsupervised approaches can help practitioners to find emerging trends or new requirements within thousands of user comments.
By looking at the emerging topics, practitioners can decide which they want to further understand and explore and remove all tweets of topics they find irrelevant (like the raffle topic) \cite{Stanik:Feedback:2019} to reduce the overall amount of comments to read \cite{Pagano:App:2013, guzman2016needle}.
Further, practitioners can identify the most discussed topics and use that information for prioritizing requirements.
For example, if users frequently comment about a new feature like the visualization of the privacy policy or about a new non-functional requirement like the ecological sustainability of the provided telecommunication services our unsupervised approach can find that information. 
Practitioners can then react accordingly, e.g., if the discussion involves a critical mass of users.
However, our approach is not designed to capture unique ideas as we configured it to extract topics that contain at least a certain number of tweets, which we set to 30 in this work.

\begin{figure}[tb]
    \centering
    \includegraphics[width=0.48\textwidth]{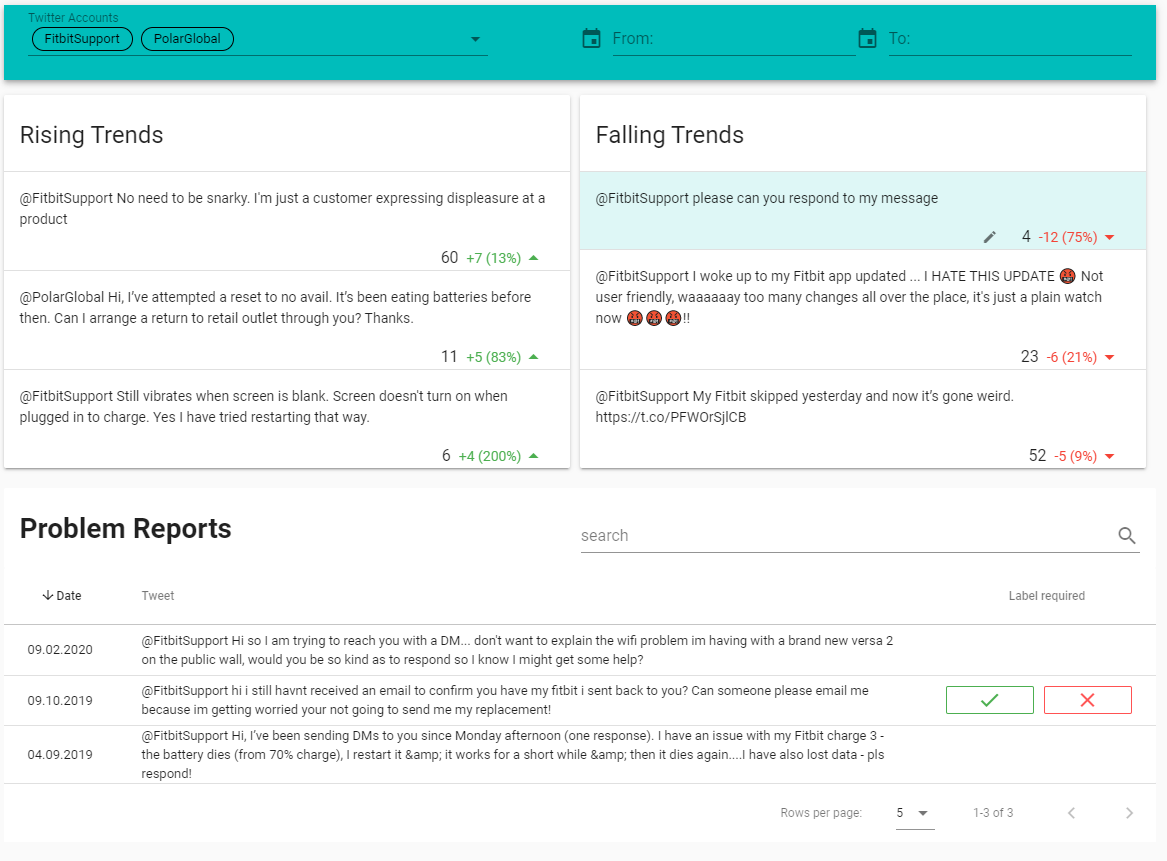}
    \caption{Integration of our approach into an open-source analytics tool.\protect\footnotemark}
    \label{fig:tool_demo}
\end{figure}

Based on the ideas and tools introduced by related work and to demonstrate how our approach can be employed in practice, we cloned the OPENREQ repository and extended its user interface.
Figure \ref{fig:tool_demo} shows our approach integration into this open source tool.
The tool already came with the filter bar on top and the table of the ``Problem Reports'' on the bottom. 
It crawls tweets continuously and classifies them into either problem reports, feature requests, or irrelevant.
We extended it with our approach by identifying trends in the problem reports.
The left side shows ``Rising Trends'', while the right side shows ``Falling Trends'' in the user comments.
As a representative of a topic, we selected the tweet that has the highest cluster assignment probability.
When selecting a tweet representative (see the blue highlighted tweet in the falling trends), the user can interact with the pen icon to see more representative tweets from the topic and to name it.

This tool integration allows practitioners to explore how many users reported which problems and if they are either new or fading away, e.g., due to bug fixes in a new version. The same analysis also applies for feature requests. 

\footnotetext{\url{https://github.com/ietz/ri-visualization}}

\subsection{Limitations and Threats to Validity} \label{sec:discussion:limitations}
\textbf{Internal.}
Evaluating the performance of clustering approaches is not an easy task even if statistical metrics indicate a high cluster cohesiveness during the configuration, as we do not know if these clusters and the resulting topics are also understandable by humans.
Therefore, we reported on a multi-step evaluation inspired by techniques from topic modeling and brought together different perspectives for human understanding of the topics.
We randomly selected our evaluation data for all evaluation steps and peer-coded all annotation tasks to reduce human bias.
For the thematic analysis, we tried mitigating potential threats to validity by ensuring the trustworthiness criteria suggested by Nowell et al. \cite{nowell2017thematic:trust}.
In particular, we involved two annotators in every step and documented our coding guide and all intermediate annotation results.
Nevertheless, potential human errors remain possible. 

\textbf{External.}
While we cited evidence that practitioners generally long for an automated solution for aggregating semantic similar user comments, we evaluated the approach for one particular large multi-national company in the telecommunication domain.
Practitioners can use our approach including its configuration and apply it to other Twitter accounts or even other user comment sources. However, we only empirically evaluated the result for one Twitter account with around 240,000 followers.
As Figure \ref{fig:tool_demo} shows, we also applied the approach to other accounts for which we had promising preliminary non-systematic assessments.
However, we did not empirically evaluate the results for multiple accounts due to the large workload. 
We decided to make an elaborate and rigorous multi-step qualitative evaluation to get a reliable idea about the approach's performance. 
By sharing the replication package and extending an open-source tool, we hope for follow-up evaluations in other domains and based on other data sources.
\section{Related Work} \label{sec:related_work}
About a decade ago, researchers identified that analyzing user comments is necessary for stakeholders to capture user requirements and to increase user satisfaction.
This section summarizes the related work regarding user comment analysis approaches and in particular, clustering and topic modeling techniques applied to user comments.

\subsection{User Comments Analytics} \label{sec:related_work:user_feedback}
Two of the most studied platforms for analyzing user comments are app stores and Twitter.
In 2017, Martin et al. \cite{Martin:TSE:2016} published an extensive survey of the field of app store analysis for software engineering.
The authors show that they can utilize app store data to develop requirements engineering, release planning, software design, security, and testing approaches.
App vendors regularly receive a large number of user comments via app stores and social media platforms as Twitter \cite{Pagano:App:2013, guzman2016needle}.
Manually filtering and processing such comments is challenging.
Therefore, research developed approaches for filtering comments, e.g., by automatically identifying relevant user comments \cite{GalvisCarreno:ICSE:2013} like bug reports \cite{Stanik:Feedback:2019, van2020identifying} and feature requests \cite{iacob2013retrieving}, or by clustering the comments \cite{Villarroel:Release:2016} to understand how many users address similar topics \cite{williams2017mining}.
Other approaches like RE-SWOT from Dalpiaz and Parente identify requirements from app competitors \cite{dalpiaz2019re}.

\subsection{Clustering and Topic Modeling on User Comments}  \label{sec:related_work:clustering}
In 2013, Carre\~{n}o Kristina Winbladh \cite{GalvisCarreno:ICSE:2013} introduce an approach based on the Aspect and Sentiment Unification Model (ASUM), a Latent Dirichlet Allocation (LDA) extension considering sentiment.
They state that ASUM requires a large number of clusters to achieve good results.
Many clusters would result in more work for stakeholders to grasp them.
Further, the approach needs manual tuning regarding the optimal number of clusters.

Iacob et al.~\cite{iacob2013retrieving} used LDA to learn about topics in feature requests, which are a collection on \textit{n} words per topic.
The disadvantages of LDA are that two main parameters need configuration---the number of topics and the number of words per topic.
The second disadvantage is that stakeholders must understand the meaning of the topic themselves by interpreting the \textit{n} words per topic.

Chen et al.~\cite{Chen:2014:AMI} present AR-Miner, a coherent approach that introduces a user comment analysis workflow including the steps preprocessing, filtering, grouping, ranking, and visualizing user comments.
In the grouping step, they combine k-means clustering with LDA topic modeling as they argue that clustering assumes that one user comment can only belong to one cluster while it could contain multiple topics.
Both techniques need manual fine-tuning to lead to potentially interpretable results.

Guzman, Ibrahim, and Glinz \cite{guzman2017little} preprocess tweets, classify them into their specified categories, group semantically similar tweets, and eventually present a weighted function to rank tweets by their relevance.
Their grouping step is the closest to our approach.
In particular, they applied topic modeling with LDA and Biterm Topic Model (BTM) \cite{yan2013biterm} and evaluated both with the word intrusion and topic intrusion method we introduced in Section \ref{sec:study_design}.
They found that BTM is superior in most cases, most likely because it was designed for short texts.
In contrast, our approach uses deep bi-directional embeddings and employs a clustering approach that automatically finds the optimal number of topics.
We further added an evaluation step for analyzing the content of the topics.
Our work's focus and contribution lie in making a rigorous evaluation of one approach that does not need a complex configuration and evidently produces human-interpretable results.
\section{Conclusion} \label{sec:conclusion}
This work addresses the issue of continuously receiving large amounts of partly redundant and similar user comments.
As a manual analysis of user comments is impractical with popular apps and services, practitioners seek automated tool support to make sense of the comments.
Research developed several classification approaches for identifying feature requests and problem reports in user comments.
However, identifying redundant, similar, and related comments, e.g., to get more details about a reported issue or assess its popularity, remains a major challenge.

To solve this challenge, we introduce an approach based on the deep bi-directional natural language understanding of SBERT. 
SBERT's underlying technology BERT, can create contextualized vectors that enable text embeddings (vectors) with richer information than previous approaches like TF/IDF or word2vec.
We then performed dimension reduction on the SBERT embeddings to prepare them for the clustering step.
We chose the HDBSCAN clustering technique as it does not require a manual step of pre-defining the numbers of clusters compared to other popular techniques like k-means.

We then performed a multi-step evaluation of the discovered topics using a fairly large Twitter dataset from the telecommunication domain.
Our evaluation goal was to check if humans can understand the extracted topics and the topics could help in requirements engineering.
In the first evaluation step (intruder detection), we evaluated the topics' semantic cohesiveness with 200 peer-coding tasks.  
We found that the annotators agreed in more than 95\% of the cases.
In a second step (document to topic assignment), two annotators reached a 98.4\% inter-coder agreement after 2,000 annotation tasks.
In a third step (expert labels), the annotators peer-coded 800 free text annotations to evaluate if both have the same understanding of the topics.
They reached an overall agreement of 91\%.
Based on the multi-step peer-coded evaluation, we have high confidence that humans can understand the topics and distinguish them when compared to other topics.

Eventually, we integrated our approach in an open-source tool for continuously analyzing tweets and illustrated how it could help practitioners identify similar problem reports and features request and grasp the frequencies and trends of comments topics.

\bibliographystyle{abbrv}
\bibliography{lib}
\end{document}